\documentclass[aps,twocolumn,showpacs,showkeys,floatfix,longbibliography,superscriptaddress]{revtex4-1}

\usepackage{graphicx}
\usepackage{textcomp}
\usepackage{dcolumn}
\usepackage{amsmath}
\usepackage{amssymb}
\usepackage{epstopdf}
\usepackage{rotating}
\usepackage{color}
\usepackage{natbib}
\usepackage{url}
\setcitestyle{square} 
\begin{document}
 
\title{Student Outcomes Across Collaborative-Learning Environments}

 
\author{Xochith Herrera}
\author{Jayson Nissen}
\author{Ben Van Dusen}
\affiliation{Department of Science Education, California State University Chico, Chico, CA, 95929, USA} 
 
\begin{abstract}
The Learning Assistant (LA) model supports instructors in implementing research-based teaching practices in their own courses. In the LA model undergraduate students are hired to help facilitate research-based collaborative-learning activities.Using the Learning About STEM Student Outcomes (LASSO) database, we examined student learning from 112 first-semester physics courses that used either lecture-based instruction, collaborative instruction without LAs, or LA supported instruction. We measured student learning using 5959 students' responses on the Force and Motion Conceptual Evaluation (FMCE) or Force Concept Inventory (FCI). Results from Hierarchical Linear Models (HLM) indicated that LA supported courses had higher posttest scores than collaborative courses without LAs and that LA supported courses that used LAs in laboratory and recitation had higher posttest scores than those that used LAs in lecture. 

 \end{abstract}
\maketitle
\section{Introduction}
A central goal of discipline-based education research (DBER) is to identify pedagogical practices that improve student outcomes \citep{national2012discipline}. A common way that DBER researchers investigate student outcomes is by comparing student performance on research-based assessments (RBA), such as the Force Concept Inventory (FCI) \citep{hestenes1992force}, between courses that use different pedagogical practices. For example, Hake's seminal investigation of student learning \citep{hake1998interactive} used pre- and posttests to find that courses that used interactive engagement had normalized learning gains approximately twice as large as those in lecture-based courses.

\par One strategy for increasing the use of research-based teaching practices is the Learning Assistants (LA) model. The LA model originated from the University of Colorado Boulder with the intention of addressing national challenges in education that engage both science and education faculty by supporting the adoption of existing or generation of new research-based pedagogical strategies \citep{otero2006responsible}. In the LA model, knowledgeable undergraduate students (LAs) work directly with faculty to support collaborative learning in undergraduate STEM courses. Learning assistants couple interactive engagement with collaborative work by facilitating small-group discussions. Instructors meet with LAs in weekly preparation meetings where they work through the content to be covered,  discuss students’ progress, and reflect on the teaching from the previous week. Since its development, the LA model has spread to over 75 institutions. 

\par The implementation of LAs has been associated with improved student outcomes \citep{otero2006responsible,white2016impacts,pollock2008sustaining}.  These studies, however, were primarily not designed to isolate the impact of LAs from the research-based pedagogical practices. It is unknown if the use of LAs increases student outcomes beyond the increase due to the research-based practices they support. 

\par Previous research shows that student outcomes are improved in courses that use collaborative learning than in traditional courses, but student outcomes vary substantially across educational environments. For example, Pollock and colleagues \cite{pollock2008sustaining} examined of the impact of LAs and research-based tutorials across two types of courses: IE lecture with traditional recitations and IE lectures with the use of LAs and tutorials. The results indicated that courses with LAs and tutorials led to more student learning, but they also showed wide variation in student outcomes across the LA-supported courses \citep{pollock2008sustaining}.
\par  \citet{white2016impacts} found meaningful variations in student outcomes across different uses of LAs. White used Cohen's d effect size to compare students’ pre- and posttest scores as a measure of student learning. They found that effect sizes were higher in LA-supported classroom contexts such as labs (1.9 times higher), lectures (1.4 times higher), and recitations (1.5 times higher) when compared to non-LA supported courses. However, in this investigation, student outcomes from LA supported courses were not compared to those from collaborative courses without LAs.

\par Although these studies associate LAs with improved student learning, there is limited research that compares student outcomes from other research-based pedagogical practices to those from LA-supported courses. Our purpose is to inform the extent to which the use of LAs improves student outcomes beyond the improvement caused by the implementation of collaborative learning that increases student outcomes. We do this by differentiating student outcomes from traditional instruction, collaborative instruction without LAs, and LA supported instruction.

\begin{table*}
\caption{Data breakdown by instrument, course type, and primary LA context.}
\begin{tabular}{lcp{.1cm}ccp{.1cm}cccp{.1cm}ccc}
\hline \hline
\rule{0pt}{2ex}    
&Total&&\multicolumn{2}{c}{Instrument}&&\multicolumn{3}{c}{Course Type}&&\multicolumn{3}{c}{Primary LA Context}\\ \cline{2-2}\cline{4-5} \cline{7-9} \cline{11-13}
\rule{0pt}{2ex}    
		&		&&FCI	&FMCE	&&Trad.	&Collab.	&LAs	&&Lec.	&Rec.	&Lab.\\ 
Courses	&112	&&92	&20		&&18	&24			&70		&&55		&9		&6\\
Students&5959	&&4077	&1882	&&791	&1068		&4100	&&2792	&1113	&195\\
\hline \hline
\end{tabular}
\end{table*}

\section{Research Questions}
\par To isolate the impact of students spending time in collaborative learning activities, we examined first semester physics courses with different teaching techniques: traditional, collaborative without LAs, and LA-supported. We further examined the association between student outcomes and uses of LAs in different classroom contexts (lab, lecture, and recitations) to identify the environments where using LAs is associated with a larger improvement in student learning. We hypothesize that structural features of classroom contexts create more opportunities for student collaboration. Labs and recitations are typically designed with low instructor to student ratios where it is easy for the instructors to move between groups. In contrast, lectures typically have high instructor to student ratios and instructor movement is restricted by rows of seats. To do this, we investigated the following research questions in introductory physics courses: 
\par (1) \emph{To what extent is the inclusion of LAs in collaborative learning environments associated with improved shifts in student outcomes?}
\par (2) \emph{To what extent do student outcomes vary across the use of LAs in lab, lecture, and recitation?}

\section{Methods}

\begin{figure*}
\centering
\includegraphics[width=2\columnwidth]{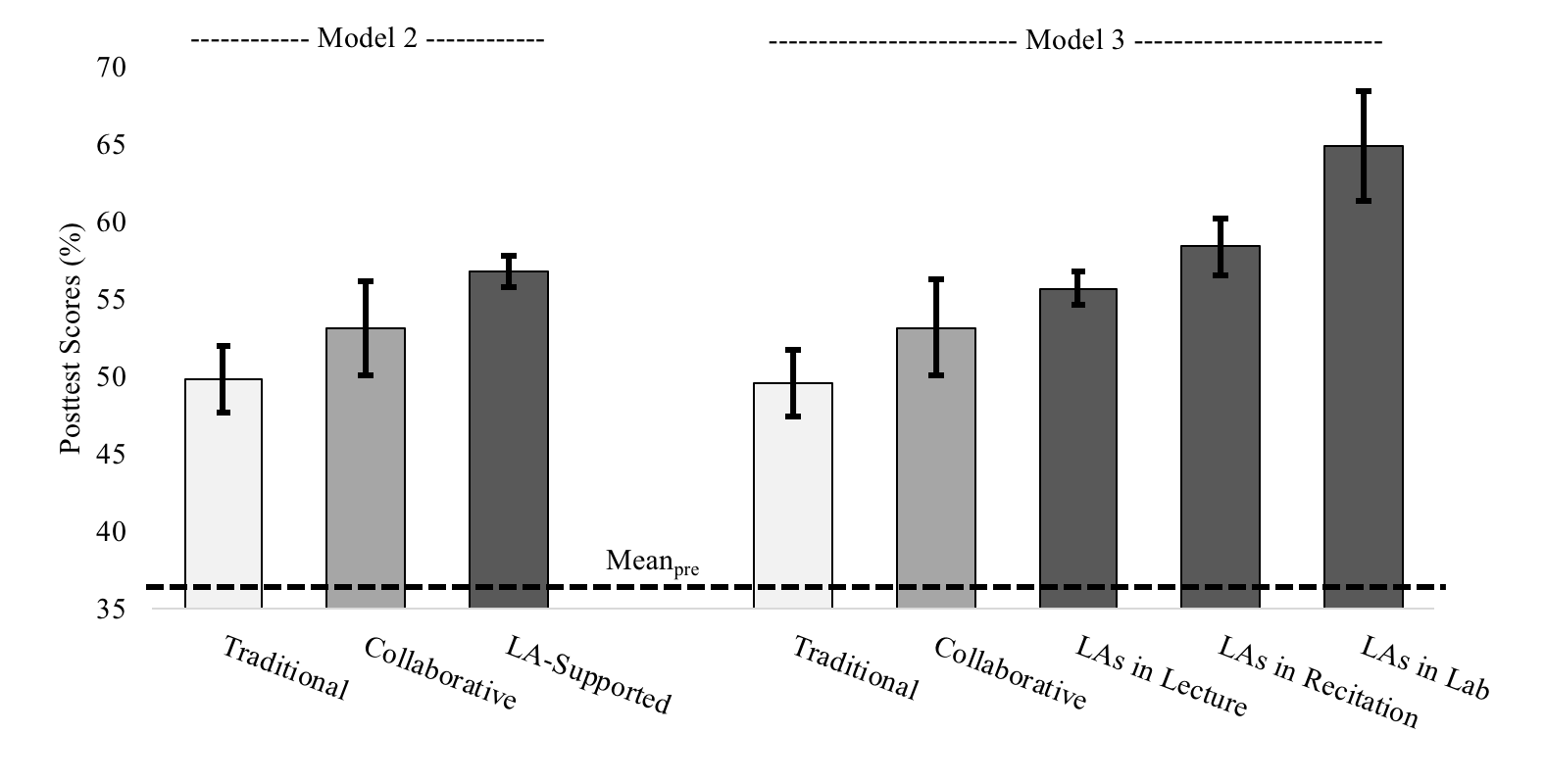}
\caption{Predicted posttest scores for traditional, collaborative, and LA supported courses. Error bars were calculated using the hypothesis testing function on HLM and are one standard deviation from the mean. Class mean pre score is 36\%.}
\end{figure*}

\par This study used three years (Spring  2015 - Summer  2017) of introductory physics course data from the LASSO database. The LASSO platform hosts, administers, scores, and analyzes student pre- and post tests online to provide instructors with objective feedback on student learning in their courses. When creating a course on LASSO, instructors are asked questions about the course. For example, ``Do students in your class engage in collaborative learning?'' and ``What is the primary use of LAs in your course?'' Instructors receive a report on their class’s performance and can access complete data for their course. The LASSO database records student responses to the RBAs and course data. The data are anonymized, aggregated, and made available to researchers with approved IRB protocols. We examined data from courses that used the FCI or FMCE. We did not differentiate between the FCI and FMCE in the models presented in this paper because our preliminary analysis indicated that doing so did not meaningfully change our models.

\par To clean our data, we removed assessment scores for students if they took less than 5 minutes to complete the assessment or if they completed less than 80\% of the questions. We removed entire courses if they had less than 40\% student participation on either the pre- or posttest. After filtering, our data was missing 15\% of the pretest scores and 30\% of the posttest scores. We used hierarchical multiple imputation (HMI) with the HMI and mice packages in R to address missing data. HMI is preferred method of cleaning data, in comparison to using matched data, because it maximizes statistical power by using all available data. HMI addresses missing data by (1) imputing each missing data point m times to create \textit{m} complete datasets, (2) independently analyzing each data set, and (3) combining the \textit{m} results using standardized methods \citep{drechsler2015multiple}. The analysis used 10 imputed datasets. After filtering and imputation, our dataset included pre- and posttest scores for 5959 students from 112 courses (Table 1).

\par We investigated our research questions using 2-level hierarchical linear models (HLM) \citep{astin2009multi} with the HLM 7.02 software. By nesting student data (level 1) within course data (level 2), these models quantify the relationships between collaborative learning strategies in courses with and without LAs and student learning. Our final model examined the differences in posttest scores in courses that implemented LAs in different classroom contexts (lecture, recitation, and labs).	
\par \textbf{Level-1 Equation}
\begin{eqnarray*} 
Post Score_{ij} & = & \beta_{0j} + \beta_{1j}*(Student Pre Score_{ij}) +r_{ij}
 \end{eqnarray*}
\par \textbf{Level-2 Equations}
\begin{eqnarray*}
\beta_{0j}& = & \gamma_{00}+\gamma_{01}*(Class Mean Prescore_{j}) +\\ 
& & \gamma_{02}*(Collaborative Learning_{j})+\\
& & \gamma_{03}*(LAs In Lecture_{j})+\\
& &  \gamma_{04}*(LAs In Recitations_{j})+u_{0j}\\ 
& &  \gamma_{05}*(LAs In Labs_{j})+u_{0j}\\ 
\beta_{1j}& = & \gamma_{10}+u_{1j}\\
\end{eqnarray*}
\par We developed three models in incremental steps. For each model, the outcome variable, was student's posttest score. Model 1 is an unconditional model, which predicts student posttest scores without level-1 or level-2 predictor variables. Model 2 builds on model 1 by integrating course (level-2) variables (class mean prescore, collaborative learning, LA supported). Model 3 builds on model 2 by including the course (level-2) variables (class mean prescore, collaborative learning, LAs in lecture, recitations, and labs). The equation for model 3 is shown below. For ease of interpretation, student prescore is group mean centered, class mean prescore is grand mean centered, and all other variables are uncentered. These centerings simplify interpreting the model by shifting the model to predict posttest scores for average performing students in average performing classes. We included pretest scores in the model because they are strong predictors of student performance and improved the model’s fit. The intercept in models 2 and 3 predicts the posttest score of a student with an average pretest score in a traditional course with an average class mean pretest score. Each coefficient in level 1 has an associated level 2 equation. In the level 2 equation, the intercept is $\gamma_{00}$, there is an associated coefficient ($\gamma_{ij}$) for each variable in the equation and $u_{ij}$ represents the random effect for the level 2 equations.

\section{Findings}
\begin{table}
\caption{Hierarchical Linear Models.}
\begin{tabular}{p{2cm}ccp{.05cm}ccp{.05cm}cc}
\hline \hline

\rule{0pt}{3ex}&\multicolumn{8}{c}{Fixed Effects with Robust SE}\\ \cline{2-9}
\rule{0pt}{3ex} &\multicolumn{2}{c}{\underline{Model 1}}&&\multicolumn{2}{c}{\underline{Model 2}}&&\multicolumn{2}{c}{\underline{Model 3}}\\
			&$\beta$&\emph{p}&&$\beta$&\emph{p}&&$\beta$&\emph{p}\\

For Intercept	&&&&&&&&\\
~~Intercept		&54.55&\textless0.001	&&49.59&\textless0.001	&&49.85&\textless0.001	\\
~~Class Pre		&-&-&	&0.92&\textless0.001 	&&0.96&\textless0.001\\
~~Collab.	&-&-&	&3.59&0.246 &&3.31&0.279\\
~~LA Sup  &-&-&	&7.24&0.003       &&-&-\\
~~LAs in Lab&-&-&			&-&-&			&15.08 &\textless0.001\\
~~LAs in Rec. &-&-&			&-&-&			&8.60&0.003\\
~~LAs in Lec. &-&-&			&-&-&			&5.85 &0.018\\

\multicolumn{3}{l}{Student Pre}&&&&&&\\
~~Intercept		&-&-&		&0.57&\textless0.001	&&0.57&\textless0.001	\\
\hline

\hline \hline

\end{tabular}
\end{table}

\par Model 1 shows that the average posttest score across all classes was 54.6\%. Given that the mean pretest score was 36.4\%, the predicted mean gain across all course types was 18.2\%. Model 2 builds on this by including variables for collaborative learning and LA-supported courses (Table 1). It allows us to predict the the posttest scores for average students in traditional (49.6\%), collaborative learning (53.2\%), and LA-supported (56.8\%) courses (Figure 1). The model shows that courses that used either collaborative learning or LA support had meaningfully reliably higher posttest scores over traditional courses (3.6\% and 7.2\%).

\par Model 3 disaggregates the uses of LAs between LAs used in lecture, recitation, and lab (Table 1). The predicted posttest scores are meaningfully and reliably larger for each LA use (55.7\% in lecture, 58.5\% in recitation, and 64.9\% in lab) than in traditional courses (Figure 1). The raw gains associated with each LA vary significantly across LA uses (from 5.9\% in lecture to 15.1\% in labs) but are all larger than the raw gain associated with collaborative learning (3.3\%). These gains align with our a priori categorization of learning environments as more or less supportive of collaborative learning.

\section{Discussion}
\par Courses that used collaborative learning are associated with more student learning than traditional courses. Traditional courses had gains from pretest to posttest of 13.2\%. Students learned 1.27 times more in courses with collaborative learning and 1.55 times more in LA supported courses than students from traditional courses. These results align with the well-established findings that courses using collaborative learning are more effective than traditional, lecture-based courses \citep{hake1998interactive}, but extend them to show that the inclusion of LAs in courses with collaborative learning is associated with additional improvement in student learning. 
\par While student post-test scores were higher in LA-supported courses, our models show variation in post-test scores across the three uses of LAs (lecture, recitation, and labs). Student post-test score were 1.46 times higher in courses that used LAs in lecture, 1.67 times higher in courses that used LAs in recitation, and 2.16 times higher in courses that used LAs in lab compared to student posttest scores from traditional instruction. The spread of gains across LA-supported courses suggest that differences in learning environments or how instructors implement LAs may have meaningful impacts on student outcomes.
\par Classroom affordances and constraints may limit the impact of collaborative learning activities. Lecture halls generally have fixed seating that face forward, not toward each other, which may limit the student to student engagement, small group work, and student interactions with instructors and LAs. On the other hand, lab settings generally consist of lab benches where small groups work together facing each other or shared equipment that may enable collaborative learning. Lab settings often have lower student to instructor (including LAs) ratios than lecture settings, potentially increasing the effectiveness of collaborative learning in them. Nonetheless, more student learning occurs in courses that use LAs in lecture than courses with traditional instruction.

\section {Limitations and future work}
\par Our findings lead us to recommend using LAs in environments that are more conducive to small group collaborative work, such as labs and recitation. Instructors using LAs in lecture may improve student outcomes by transforming the lecture environment to better support small-group collaborative activities. For example, instructors may arrange desks in small groups or in a lecture hall they may ask students to sit in every other row so that LAs and instructors can more easily interact with them. 
\par Our study is exploratory and cannot definitively identify the cause of higher outcomes in collaborative courses with LAs over those without LAs. For example, there may be differences between the instructors or the institutions that use LAs and those who don't that impact student outcomes. Instructors that choose to use LAs may be actively searching for ways to improve their student learning. The use of LAs may correlate with, but not cause, increased time spent in collaborative learning activities. Student outcomes may also be affected by differences in institutions. Students with better prior preparation in math are more likely to attend well funded institutions which, in turn, are more likely to be able to support the cost to implement LA programs.
\par Future research may further account for variations in student outcomes across learning environments by developing models that include student, instructor and institutional variables. Promising variables include students’ socioeconomic status, prior achievements, genders, race/ethnicity, teachers’ experiences and attitudes about teaching, and institutions’ Carnegie Classification. The present work accounts for some of this variation by controlling for pretest scores at both the student and course level.

\section {Acknowledgements}
\par This work is funded in part by NSF-IUSE Grant No. DUE-1525338 and is Contribution No. LAA- 054 of the Learning Assistant Alliance. We are grateful to the Learning Assistant Program at the University of Colorado Boulder for establishing the foundation for LASSO and LASSO studies.
\bibliography{sample}

\end{document}